# Monte-Carlo simulations of the expected imaging performance of the EXIST high-energy telescope

S. V. Vadawale[*a], J. Hong[a] and J. E. Grindlay[a] and G. Skinner[b]

[a]Harvard-Smithsonian Center for Astrophysics, Cambridge, MA, USA
[b]CESR, Toulouse, France

## Abstract

EXIST is being studied as the Black Hole Finder Probe, one of the 3 Einstein Probe missions under NASA's Beyond Einstein program. The major science goals for EXIST include highly sensitive full-sky hard X-ray survey in a very wide energy band of 5 - 600 keV. The scientific requirements of wide energy band (10-600 keV for the High Energy Telescope considered for EXIST) and large field of view (approximately 130º x 60º in the current design, incorporating an array of 18 contiguous very large area coded aperture telescopes) presents significant imaging challenges. The requirement of achieving high imaging sensitivity puts stringent limits on the uniformity and knowledge of systematics for the detector plane. In order to accomplish the ambitious scientific requirements of EXIST, it is necessary to implement many novel techniques. Here we present the initial results of our extensive Monte-Carlo simulations of coded mask imaging for EXIST to estimate the performance degradation due to various factors affecting the imaging such as the non-ideal detector plane and bright partially coded sources.

## 1. Introduction

**1.1 Black Hole Finder Probe**

The Energetic X-ray Imaging Survey Telescope[1] (EXIST) is one of the two mission concepts for the Black Hole Finder Probe (BHFP) selected for further study by NASA. BHFP forms a very important part of the NASA's future program "Beyond Einstein"[2]. The major goal for the BHFP is to perform a census of black-holes in the universe on all scales i.e. from stellar mass black holes to super-massive black holes, which are expected be achieved best by an all-sky survey in the hard X-ray band (5-600 keV). With the EXIST concept, this will be achieved by surveying the entire sky every orbit with the limiting sensitivity of ~50 µCrab (10-150 keV) in one year of survey time. EXIST will also serve as a next-generation GRB mission and will be able to detect GRBs up to the redshift of z=15, thus seeing the black holes formed by the death of the very first stars (population III stars) formed in the universe. Other major science goals for EXIST are: (a) estimate the total accretion luminosity of the local universe by determining the fraction of obscured AGNs; (b) improve our understanding of physical processes occurring in the extreme environments of high gravity, high magnetic field, and high radiation energy by long duration timing studies of accreting X-ray binaries; (c) find the hidden supernovae and novae by detecting the characteristic nuclear lines and thus provide the crucial information on supernova and nova rates in the galaxy to constrain the models of cosmic nucleosynthesis; (d) understand the formation and evolution of neutron stars with the most extreme magnetic fields in the local universe through the high-sensitivity studies of soft gamma-ray repeaters. It should be noted that the sky has not been surveyed in the hard X-ray band during last 25 years, the last (and only true all-sky) survey being HEAO1-A4 survey[3] with the sensitivity ~50 mCrab. Thus EXIST will be the first truly deep all-sky hard X-ray survey with sensitivity comparable to that in the soft X-ray band (ROSAT all sky survey[4] with 50 µCrab sensitivity), and hence it is anticipated that it will also bring many unexpected discoveries.

---

* Send correspondence to S. V. Vadawale (*svadawale@cfa.harvard.edu*)

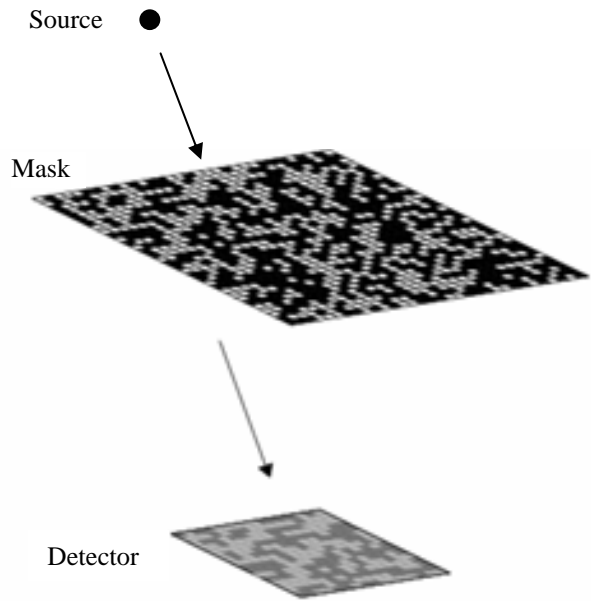

**Figure 1.** Basic principle of coded aperture imaging. A position sensitive detector records a shadowgram of the coded mask. Knowledge of the coding pattern is then used to reconstruct the source image.

**1.2 Coded Aperture Imaging**

The fundamental requirement for achieving all the science goals of EXIST (or BHFP in general) is high-sensitivity, high-resolution imaging in the 5 – 600 keV energy band. A typical approach to obtain high-sensitivity, high-resolution images is to use focusing optics. However, X-ray focusing optics is not suitable for wide-field hard X-ray imaging, because it typically has very small field of view and also is generally limited to energies less than 100 keV. Thus for EXIST, it is necessary to implement some indirect imaging method, such as coded aperture imaging. Coded aperture imaging is one indirect imaging technique which is widely used for hard X-ray imaging, Swift[5] and INTEGRAL[6] being the latest examples of hard X-ray missions using coded aperture imaging. The basic principle of coded aperture imaging is described in Figure 1. It is a two-step method[7,8]. The first step involves recording the shadowgram of the coded mask (a mask of opaque and transparent elements coded in some particular pattern) illuminated by a source. In the second step, the source image is reconstructed by applying the knowledge of the coding pattern to the recorded shadowgram. The quality of the resulting image depends on the chosen mask pattern. The mask pattern is typically described by a binary array i.e. an array of 1 and 0 where 1 denotes a transparent element and 0 denotes an opaque element. In general, any binary array (even a random array of 1 and 0) can be used to construct the coded mask. However, there are some patterns, known as URA (Uniformly Redundant Array), which have optimum imaging properties for coded aperture imaging[9]. The URA is defined by the mathematical property that its periodic auto-correlation function is a delta function. A nice review by Busboom et al.[10] describes different methods for constructing URAs of various sizes. In the current baseline design, EXIST will use URA masks for its coded aperture telescopes.

**1.3 Present EXIST mission concept**

The present mission concept for EXIST is shown in Figure 2. It uses two different telescope systems, the Low Energy Telescope (LET) and the High Energy Telescope (HET) to cover the full energy band of 5 – 600 keV. Both LET and HET utilizes the coded aperture imaging technique. The LET consists of 28 individual sub-telescopes and covers the 5 – 30 keV energy range. The detector plane for LET is an array of pixilated Si detectors with a pixel pitch of about 150 µm, with the total LET detector area being approximately 1.5 m$^2$ and the total field of view of LET being comparable to that of the HET. The main goal of the LET is to localize the sources with much better accuracy than HET to enable unique galaxy-AGN identifications at the survey limit. The HET, with energy range of 10 – 600 keV, is the primary science instrument for EXIST for realizing the goals of BHFP. It consists of 18 individual sub-telescopes arranged in a 3 x 6 array. Each sub-telescope has a field of view of about 20º x 20º giving a total HET field of view of approximately 130º x 60º. The detector plane for HET is a closely tiled array of pixilated CZT detectors with a pixel pitch of about 1.25 mm, with the total HET detector area being approximately 6 m$^2$. EXIST will have a low earth orbit with approximate altitude of ~500 km (orbital period of ~95 min), and with the help of an extremely large field of view, it will survey the entire sky every orbit. With the largest total detector area, EXIST is expected to survey the hard X-ray sky with an unprecedented sensitivity of ~50 µCrab in the 5 – 150 keV energy range (and ~0.5mCrab in the 150-600 keV band), which would be at least 1000 times deeper then any previous full-sky hard X-ray survey.

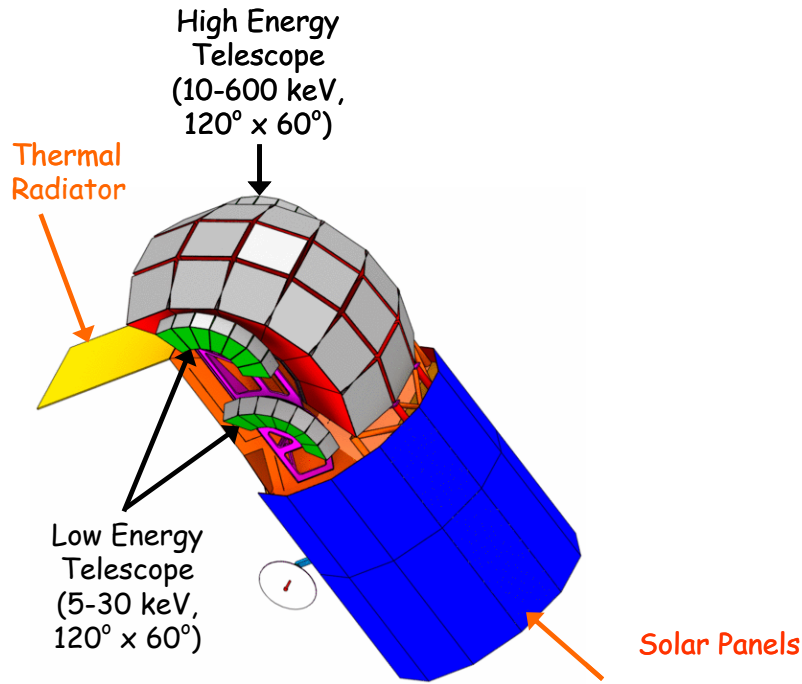

**Figure 2.** Present mission concept for EXIST

## 2. Imaging issues for EXIST

Because of its very large scale, EXIST faces some unique problems in achieving the optimum imaging sensitivity. The eventual aim of our exercise is to estimate the imaging sensitivity of EXIST considering all possible factors affecting it. In this paper, we investigate the effect of the two most critical problems – non-uniformity in the detector plane and the partially coded sources – on the imaging performance of the EXIST high energy telescope, as well as the probable strategies to minimize these adverse effects. Other EXIST Working Groups (i.e. that for detector backgrounds) are presently investigating other issues such as spatial variation of the instrument background, orbital dependence of the instrument background, background due to activation, shield leakage, etc. In our future imaging and mission simulations, we plan to include most of these effects.

**2.1 Detector plane non-uniformity and scanning**

In order to detect very faint sources, each telescope must have a very large detector plane which means that the detector plane must have a tiled array of smaller crystals. Tiling multiple crystals inevitably leaves gaps between crystals. Such gaps might range from approximately 100 µm which would be the optimum packaging limit, to a few mm if required by the electronics readout or mechanical packaging. Similarly, to achieve the fine angular resolution, the detector pixel size must be very small and thus the total number of pixels in the detector plane must be very large. In the current baseline design the total number of pixels in the full HET is expected to be approximately 3 million. Having such a large number of pixels means that there will be unknown systematic variations in the efficiency and gain among the pixels or even some dead pixels in the detector plane. The gaps between crystals are an example of a "known" systematic non-uniformity in the detector plane, whereas the efficiency variation and the dead pixels are examples of unknown (at a precise level) systematic non-uniformities in the detector plane. Either of them will have some effect on the imaging performance of the EXIST telescope.

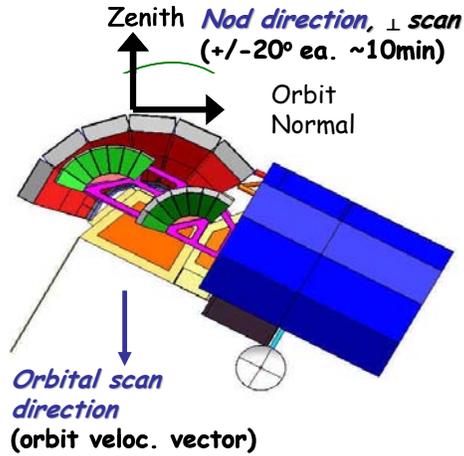

**Figure 3.** Two-dimensional scanning for EXIST, orbital velocity (in out of the page direction) provides scanning in one dimension and nodding provide scanning in the second dimension

One possible approach to overcome the problems due to non-uniformities in the detector plane is scanning [11]. In this approach, the pointing direction of the coded aperture telescope is changed continuously (or in very small steps) such that the X-ray source of interest passes across the full field of view. The final image is then obtained by co-adding the images of the individual steps. The basic idea of scanning is that the non-uniformity in the detector is exposed by different parts of the sky in the images of individual steps, and when all such images are added together, any artifacts in the individual images get averaged out from the final image. In order to take full advantage of the scanning, it is necessary to scan the complete mask pattern in both dimensions of the detector plane. One major advantage EXIST is that scanning in one dimension can be achieved very easily by utilizing the orbital motion of the satellite, and in fact is fundamental to the Survey coverage. To achieve scanning in the orthogonal direction, it is proposed that the satellite will also have a nodding motion of $\pm 20°$ with respect to the zenith (see Figure 3). Thus, the main purpose of our present exercise is to derive the performance of scanning by means of a full Monte-Carlo simulation.

**2.2 Partially coded sources and cleaning the image**

Another major imaging problem for the EXIST telescope is due to its very large field of view. The large field of view is required to achieve the whole sky survey every orbit, which results in a very large partially coded field of view. The sources in the partially coded field of view, i.e. the sources which do not cast the shadow of the full mask pattern on the detector plane, have undesirable effects on the fully coded image. Such partially coded sources generally appear in the fully-coded image as "ghost" point sources but more importantly these sources also generate coding noise which is distributed over the full image plane. The ghost sources themselves are not a big problem because they appear at a predictable location relative to the partially coded source and hence are easy to identify. However, the coding noise due to a partially coded source, particularly if the source is bright, severely degrades the sensitivity in the fully coded field of view. In the present exercise, we quantify the effects of a bright partially coded source over the fully coded image. We also demonstrate a method to eliminate the artifacts due to a partially coded source by "cleaning" the partially source contribution from the shadowgram as described in the next section.

## 3. Simulations: Method

We are developing a software tool to carry out realistic Monte-Carlo simulations of the imaging performance of the EXIST HET. The eventual aim is to simulate the all sky image using a real source catalog (such as HEAO-A4 catalog as well as the INTEGRAL and Swift high energy source catalogs) as input and considering most of the aspects (as mentioned in section 2) affecting the imaging performance. Such a tool will be very useful to get an estimate of the limiting sensitivity of any given configuration of EXIST HET and thus to optimize the optical configuration with respect to the ultimate goal of achieving the best possible sensitivity under the usual spacecraft constraints of size, mass, power etc. Here we report the initial results of our simulations using only one EXIST HET sub-telescope (i.e. 1 of the 18 telescopes of the complete HET) to estimate the effect of detector plane non-uniformities as well as partially coded sources.

**Table 1.** Parameters used for the configuration of one EXIST HET Sub-Telescope

| Detector plane size | 528 mm x 528 mm |
|---|---|
| Detector pixel pitch | 1.375 mm x 1.375 mm |
| Pixels per CZT crystal | 16 x 16 |
| Crystals per DCA[1] | 2 x 2 |
| DCA per detector | 12 x 12 |
| CZT crystal size | 22 mm x 22 mm |
| DCA size | 44 mm x 44 mm |
| Mask size | 1056 mm x 1056 mm |
| Mask pixel size | 2.75 mm x 2.75 mm |
| Mask to detector distance | 1500 mm |
| Angular resolution | 6.3' |
| Fully coded field of view | 19.96º x 19.96º |
| Partially coded field of view (zero intensity) | 55.67º x 55.67º |

1. DCA (Detector Crystal Array) is the basic building block of the detector plane consisting of the CZT crystals and front-end as well as back-end readout electronics

**3.1 Optical configuration**

The optical configuration of the single EXIST HET Sub Telescope used in our simulation is summarized in Table 1. It should be noted that the total size of the detector plane and the pixel pitch given in the Table 1 differ from those mentioned in the introduction. We use the detector plane size of 52.8 cm x 52.8 cm with pixel pitch of 1.375 mm (giving total 384 x 384 pixels) because these values are favorable for the URA mask design and compatible with the size constraints due to the diameter of the launch vehicle (probably Delta-IV medium or heavy). At this point, these values are essentially place holder values for the true detector size and pixel pitch which will get finalized in future design revisions of EXIST.

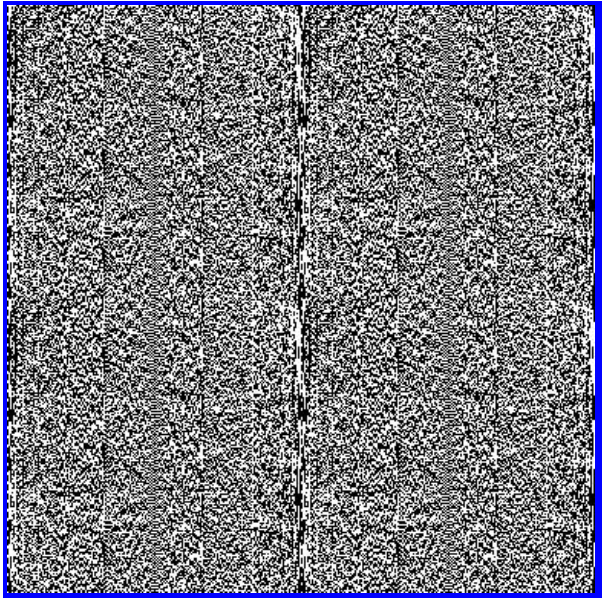

**Figure 4.** Coded mask used in our simulations. The mask pattern is 2 x 2 cyclic repetition of base PBA pattern of size 192 x 192.

For the EXIST HET mask, we use 2 x 2 cyclic repetition of the base URA pattern of size 192 x 192. The base pattern used here belongs to the class of perfect binary array (a subfamily of URA[10]. The advantage of perfect binary array is that they are available in square shape and thus makes the imaging as well as packaging design much easier. The chosen mask size of 105.6 cm x 105.6 cm and mask pixel size of 2.75 mm are twice the total detector size and detector pixel size respectively so that the coded mask is over sampled by the detector pixels by a factor of 2 (for Nyquist sampling).

**3.2 Software design**

Actual implementation of the Monte-Carlo method in our software is very simple. The basic steps involved in generating a simulated image for one HET are:

1. Set up the telescope geometry from the basic input parameters such as total number of detector and mask pixels and their size, mask detector distance, exposure time, pointing direction etc.
2. Get the list of all sources within the total field of view from the input source catalog.
3. Calculate the total number of photons incident on the coded mask from each source within the field of view using the input source flux, total mask area and total exposure time.
4. Generate the detector plane efficiency at each pixel using the input parameters such as gaps between crystals and DCAs, dead pixels, random efficiency variations etc. for "real" CZT efficiency variations.
5. Generate individual photons at a random location on the mask; test whether the photon passes through the mask; if so, track the photon up to the detector plane considering the source position; record the detector pixel in which the photon is detected considering the efficiency in that pixel.
6. Repeat step 4 for each photon for each source in the field of view
7. Calculate the total number of photons from the diffuse X-ray background (diffuse XRB flux is an input parameter); simulate the diffuse X-ray background by generating each photon with random direction within the total partially coded field of view and repeating step 4.
8. Clean the partially coded source from the shadowgram if any and if required (an input parameter)
9. Reconstruct the image by cross-correlating the shadowgram with the reconstruction array.

Cross-correlation of the shadowgram and the reconstruction array is implemented with an external FFT (Fast Fourier Transform) library "*fftw*"[12]. The reconstruction array G(x,y) is derived from the mask array M(x,y) by substituting the M(x,y) = 0 (opaque elements) G(x,y) = (1-K)/(N-K) where N is the total number of elements in the mask and K is the total open elements in the mask[13]. For these simulations with only one HET Sub Telescope, scanning is implemented by sweeping the pointing direction in an 'X' pattern (i.e. from the one corner to the opposite corner of the original FCFoV specified by the input pointing direction) in the specified number of steps (typically of the size of angular resolution). At each step during scanning the simulated image is generated for the appropriate exposure time (input exposure time divided by the number of scan steps) and these images are added together at the end.

The cleaning of the partially coded source is based on two basic assumptions: (1) the pointing direction of the telescope is precisely known and (2) the location of the partially coded source in the sky is precisely known. For the present, we envision the cleaning process as the second step in analysis and hence we assume that the flux of the partially coded source is also known. (Since for EXIST the individual scan step will be typically of the order of one second, the cleaning process requires the flux as an input parameter. In reality it will be determined for each source from its fully-coded data: each source is fully-coded for the time it transits the FCFoV.) In the actual cleaning process, the first step is to determine the pixel illumination factor (PIF) due to the partially coded source. The next step is to estimate the counts per pixel due to this source using the PIF and the source flux. The average counts per pixel, weighted by the PIF at any particular pixel, are then subtracted from the shadowgram. It can be seen that the cleaning process does leave the Poisson noise contribution of the partially coded source in the image. However, the coding noise due to the partially coded source, which is the main source of the artifact in the image, can be removed from the image. Variability of the partially coded source is also considered (see below).

## 4. Simulations: Results

Here we report the initial results of our simulations of imaging performance of one EXIST HET Sub Telescope, particularly with respect to the non-uniformities in the detector plane and the bright sources in the partially coded field of view.

**4.1 Detector non-uniformity and scanning**

Figure 5 shows three images of a bright (100 mCrab) on-axis source (1 Crab ~ 0.3 ph cm$^{-2}$ s$^{-1}$ at ~20-100 keV) with 1 day exposure time and steady pointing. The image in fig. 5a is for the ideal detector plane with no non-uniformities and thus has the best signal-to-noise ratio (SNR, defined as the total source counts divided by square root of the source + background) among the three images. For images in fig. 5b and 5c, the detector plane has 0.6 mm and 1.2 mm gaps between every CZT crystal (i.e. every 22 mm) and the SNR for both images degrades appreciably. These results are summarized in Figure 6 which shows the variation of SNR with increasing gap between crystals for the on-axis source with different intensities for 1-day exposure time. The dashed lines represent the SNR for the images with steady pointing and the solid lines represent the SNR for scanning images. It can be seen that the SNR for the images with steady pointing decrease almost linearly with the increasing gap size whereas the scanning (as described in the previous section) can recover the SNR to the almost ideal value, except for gaps larger then ~2 mm. The reason for decrease in the SNR for scanning images with very large gaps is the decrease on the total active area of the detector plane of fixed size. Figure 7 shows similar variation of SNR for pointing and scanning images with an increasing fraction of randomly distributed dead pixels in the detector plane. Again it can be seen that the scanning recovers the SNR almost up to the ideal value except for a very large fraction of dead pixels. The gaps between every crystal and the randomly distributed dead pixels are the extreme example of known vs. unknown systematic non-uniformities in the detectors. Our results show that scanning can recover the image degradation from either type of non-uniformity. However, it is still necessary to keep either type of non-uniformities to a minimum in order to maximize the total active area of the detector.

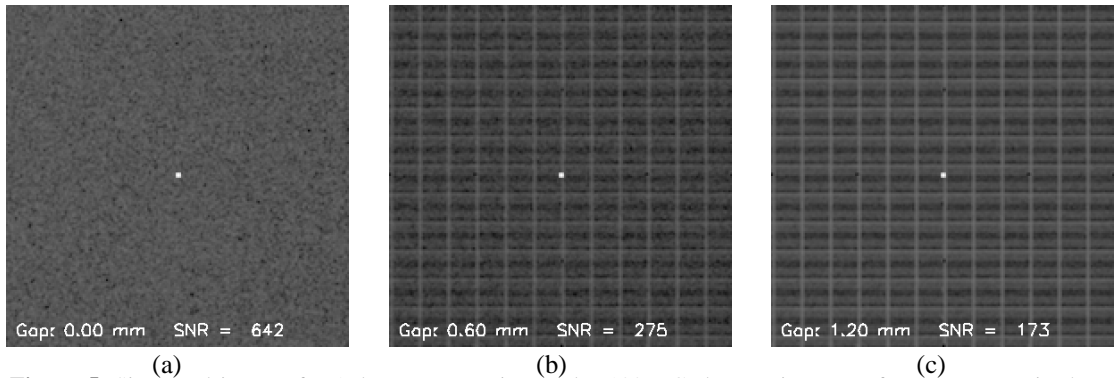

**Figure 5.** Simulated images for 1-day exposure time and a 100-mCrab on-axis source for (a) no gaps in the detector plane (b) 0.6 mm gap between every CZT crystal and (c) 1.2 mm gap between every CZT crystal. (The color-scale for these images is logarithmic to emphasize the imperfections in the images.)

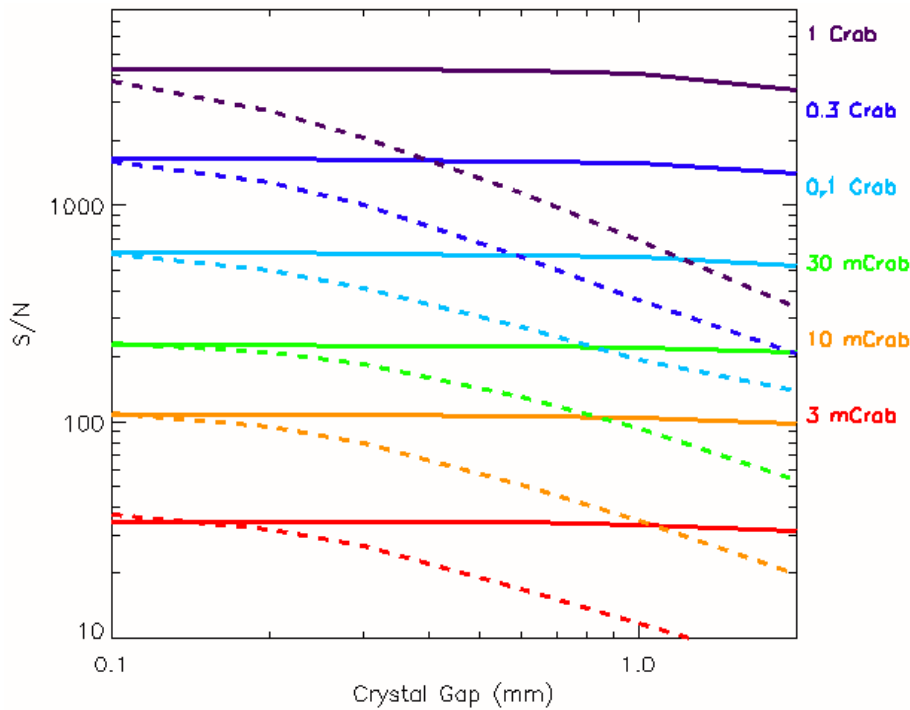

**Figure 6.** Variation of SNR with increasing gap between every crystal for 1 day Exposure time and different intensity of the on-axis source (in the order from top to bottom as given on the right side of the plot). The dashed lines represents SNR for images with steady pointing and the solid lines represent SNR for scanning images.

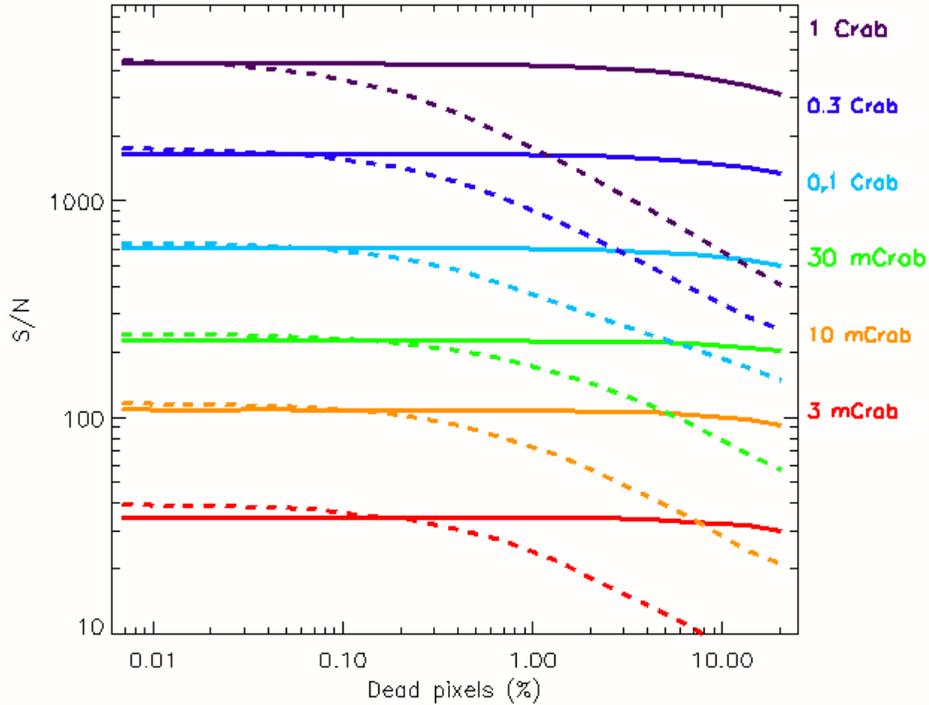

**Figure 7.** Variation of SNR with increasing percentage of dead pixels for 1 day exposure time and different intensity of the on-axis source (in the order from top to bottom as given on the right side of the plot). The dashed lines represents the SNR for images with steady pointing and the solid lines represent the SNR for scanning images.

**4.2 Partially coded sources and cleaning**

Figure 8 shows the effect of a bright partially coded source over the fully coded field of view. In the first image there are two sources, an on-axis 100 mCrab source and an off-axis 1 Crab source which is very close to the edge of the fully coded field of view. In the second and third image, the 1 Crab source is shifted away from the telescope point direction by 5 degree and 10 degree respectively, and thus this source is partially coded. The partial coding of the bright source generates a ghost source as well as a large coding noise in the fully coded field of view and hence the SNR for the on-axis source is severely degraded.

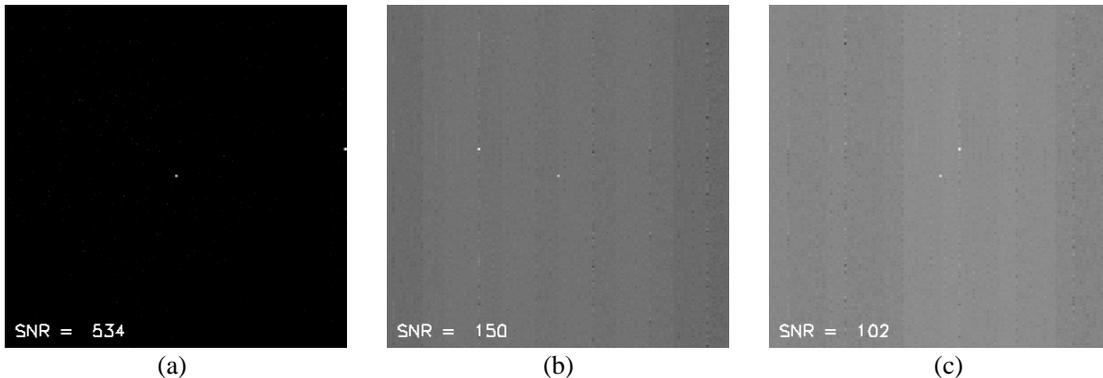

(a)          (b)          (c)

**Figure 8.** Effect of a 1-Crab partially coded source on the image of a 100-mCrab on-axis source. In the image (a) the bright source is on the right edge of the fully coded field of view and in images (b) and (c) it is shifted away by 5º and 10º from the fully coded field of view respectively. (Color-scale is logarithmic.)

However, if the contribution of the partially coded source is 'cleaned' from the shadowgram as described in the previous section then it is possible to remove both the ghost source and the coding noise in the final image (see Figure 9). The 'cleaning' procedure only subtracts the average counts per pixel and thus leaves the Poisson noise associated with the partially coded source and hence the SNR of the 'cleaned' image is not the same as the SNR for the image without the partially coded source. However, cleaning does remove the coding noise and recover the SNR up to the Poisson limit. This is summarized in Figure 10, which shows the variation of the SNR with increasing intensity of the partially coded source for 1-day exposure time. It can be seen that for the 1-day exposure the SNR drops sharply when the intensity of the partially coded source is more than ~50 mCrab. It is interesting to note that this 'cut-off' intensity of the partially coded source is independent of the intensity of the fully coded sources which suggests that the noise due to the partial coding is larger then the random diffuse noise. However, it does depend on the exposure time as is evident from Figure 11 and 12. It should be noted that the exposure times of the 1 day, 1 week and 1 month for Figures 10, 11 and 12 respectively are the actual on-source times. In terms of EXIST mission time, they correspond to 6 days, 6 weeks and 6 months respectively because of the total 60° field of view of EXIST in the direction of the orbital motion (each source in the fully coded field of view for ~16 min. during every 95 min. orbit). Thus the 'cut-off' intensity of ~5 mCrab in Figure 12 suggests that the partially coded source will not be a major problem for EXIST imaging in the high Galactic latitude sky as the bright X-ray source 3C273 in the high Galactic latitude sky has only ~5 mCrab intensity. On the other hand, in the Galactic plane there are many X-ray sources brighter then ~50 mCrab and thus the imaging in the Galactic plane will be affected by the bright partially coded sources. However, even in this case the 'cleaning' should be able to remove the effects of the bright partially coded sources and recover the sensitivity up to the Poisson limit. In general, we find that without cleaning, partially coded sources begin to degrade the fully-coded source sensitivity when the intensity of the partially coded source is brighter than about 100X the sensitivity limit for a given integration time (e.g. 5mCrab high latitude sources degrade the ultimate Survey limit of 0.05mCrab for the 1year survey).

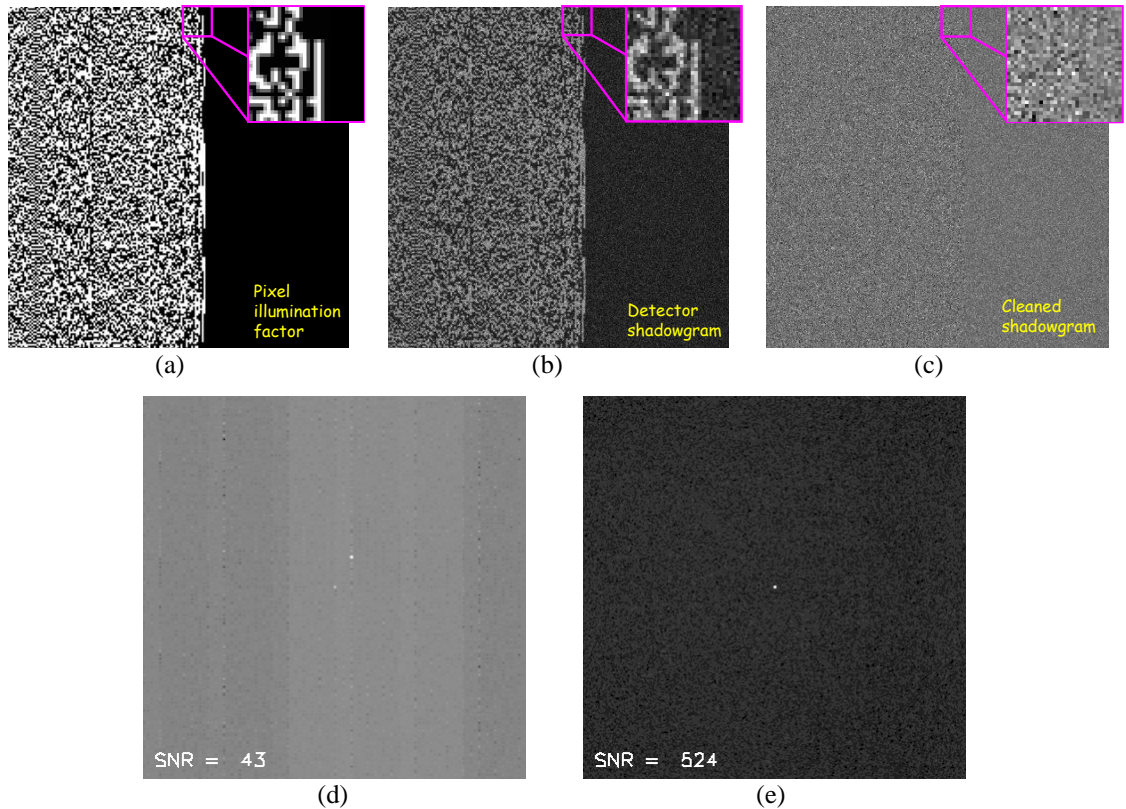

**Figure 9.** Cleaning at work – (a) pixel illumination factor for the partially coded source, (b) raw detector shadowgram (c) cleaned shadowgram, (d) uncleaned image (e) cleaned image. (Color-scale for images (d) and (e) is logarithmic.)

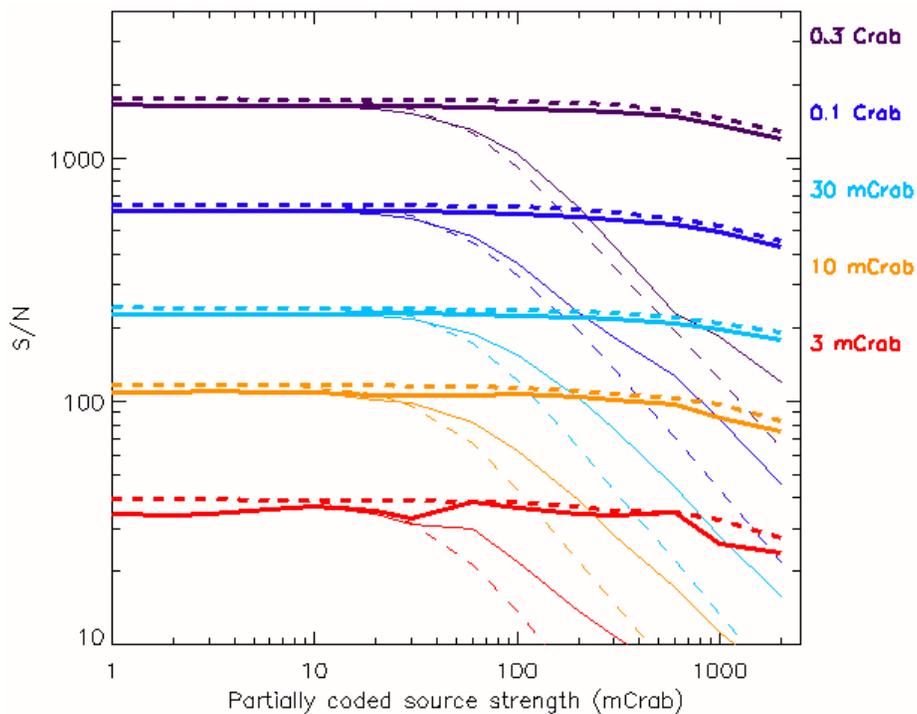

**Figure 10.** Variation of the SNR with increasing intensity of the partially coded source for a 1-day exposure time for different intensities of the on-axis source. The dashed and solid lines represent the SNR for pointing and scanning images respectively. The thick and thin lines represent the SNR for images with and without cleaning respectively.

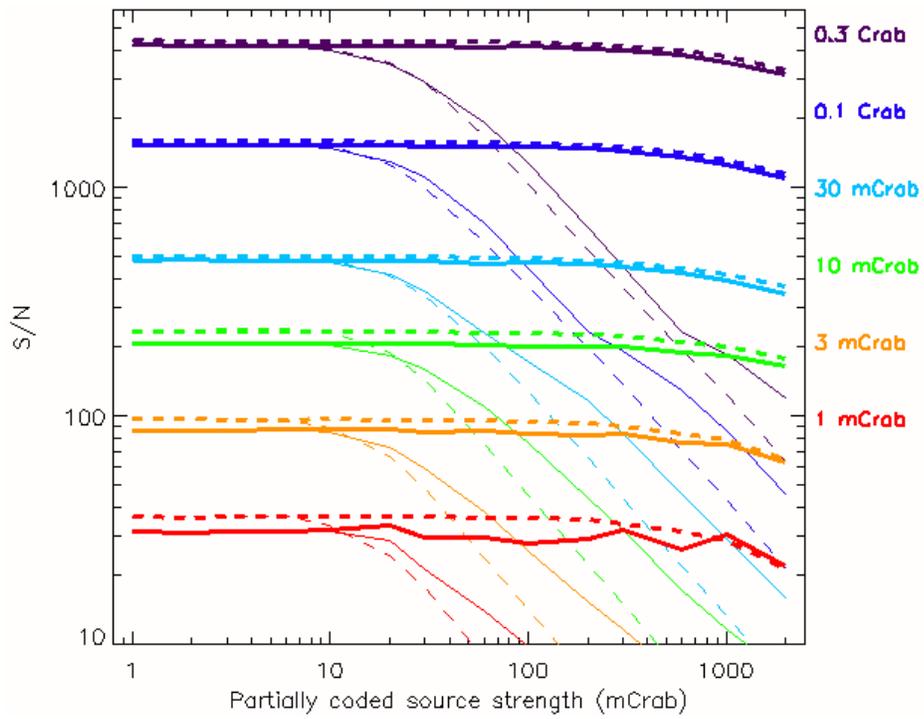

**Figure 11.** Variation of the SNR with increasing intensity of the partially coded source for a 1-week exposure time for different intensities of the on-axis source. The dashed and solid lines represent the SNR for pointing and scanning images respectively. The thick and thin lines represent the SNR for images with and without cleaning respectively.

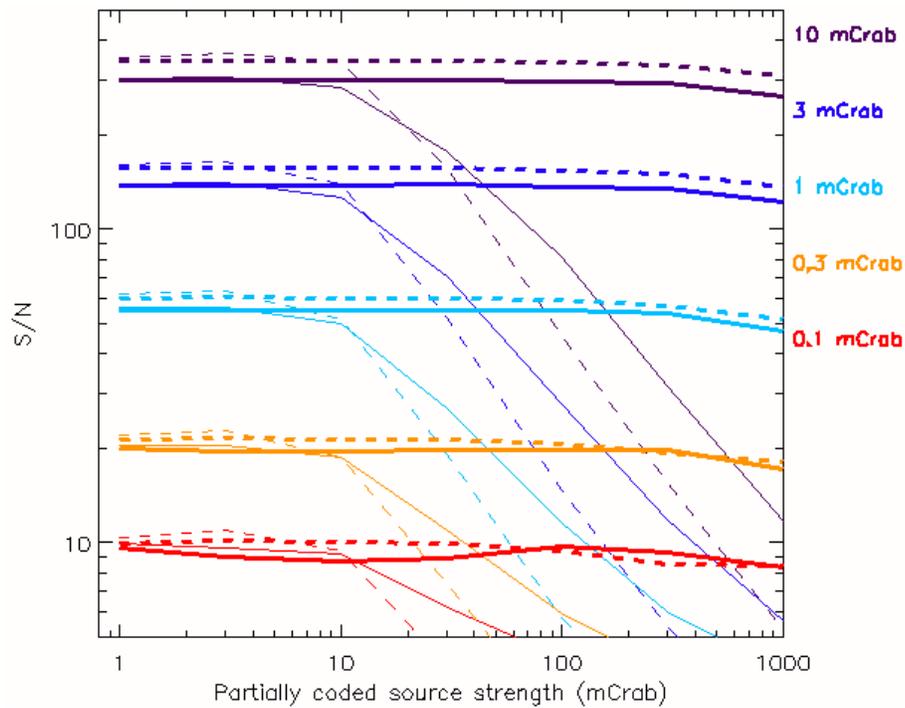

**Figure 12.** Variation of the SNR with increasing intensity of the partially coded source for a 1-month exposure time for different intensities of the on-axis source. The dashed and solid lines represent the SNR for pointing and scanning images respectively. The thick and thin lines represent the SNR for images with and without cleaning respectively.

**4.3 Temporal variability of the sources**

Our simulations so far are essentially steady state simulations. The assumptions that the sources will not vary over the time scale of a day is of course not correct. We plan to carry out full time dependent simulations to systematically study the effects of temporal variability of the source on the imaging performance. Another critical question we would like to investigate is that of the minimum time scale over which the variability to the source can be detected, given the requirement of scanning to improve the imaging, and how this minimum time scale changes with the intensity of the sources of interest. However, we did do a preliminary study of effects of source variability, particularly for the partially coded source, because the cleaning procedure is sensitive to source variability. We find that incorrect cleaning of the partially coded source does degrade the 'cleaned' SNR; however, the cleaning procedure does not break down for variability of the order of ±20 %. It should be noted that EXIST will survey the entire sky every orbit and hence it will be possible to have an accurate estimate of the flux for the bright sources every 95 minutes. One interesting point we find is that the over-subtraction of the partially coded source is more troublesome than under-subtraction, which is helpful for the cleaning of the intra-orbit variable sources, because in the case of the astrophysical source, sudden increase (i.e. under-subtraction while cleaning) is much more likely then sudden decrease (i.e. over-subtraction while cleaning). Thus, at this stage we believe that it will be possible to implement cleaning as the second step analysis procedure for the all-sky imaging with EXIST.

# 5. Conclusions

We have presented preliminary results of our simulation of the EXIST HET imaging performance and the effects of non-uniformities in the detector plane as well as of the bright partially coded sources. We find that the non-uniformities in the detector plane significantly degrade the imaging sensitivity of the coded aperture telescope with URA mask. However, we show that continuous scanning, as suggested for EXIST, does eliminate the artifacts of the detector non-uniformity and recovers the imaging sensitivity. We also find that partially coded sources brighter then ~100X the sensitivity limit for a given survey time begin to significantly degrade the imaging sensitivity, but it should be possible to restore the sensitivity using the 'cleaning' procedure suggested here.

**Acknowledgements**

We would like to thank Lars Hernquist and Suvendra Dutta of Institute for Theory and Computation (ITC) at the Harvard Smithsonian Center for Astrophysics for allowing us to use their Beowulf clusters and for help with using these clusters. The extensive simulations conducted (and planned) would not be possible without using Beowulf clusters.